\documentclass[letterpaper,twocolumn,fleqn]{article} 

\usepackage{ist}
\usepackage{amsmath}
\usepackage{amssymb}

\usepackage{soul, color}
\DeclareMathOperator*{\argmin}{argmin} 

\title{Recognition-Aware Learned Image Compression}

\author{Maxime Kawawa-Beaudan, Ryan Roggenkemper, Avideh Zakhor; Department of Electrical Engineering and Computer Science,  University of California, Berkeley; Berkeley, CA, USA}

\date{} 

\begin{document} 

\maketitle 

\thispagestyle{empty} 


\begin{abstract}
Learned image compression methods generally optimize a rate-distortion loss, trading off improvements in visual distortion for added bitrate. Increasingly, however, compressed imagery is used as an input to deep learning networks for various tasks such as classification, object detection, and super-resolution. We propose a recognition-aware learned compression method, which optimizes a rate-distortion loss alongside a task-specific loss, jointly learning compression and recognition networks. We augment a hierarchical autoencoder-based compression network with an EfficientNet recognition model and use two hyperparameters to trade off between distortion, bitrate, and recognition performance. We characterize the classification accuracy of our proposed method as a function of bitrate and find that for low bitrates our method achieves as much as 26\% higher recognition accuracy at equivalent bitrates compared to traditional methods such as Better Portable Graphics (BPG). 
\end{abstract}

\begin{figure*}[t]
  \includegraphics[keepaspectratio, width=\textwidth]{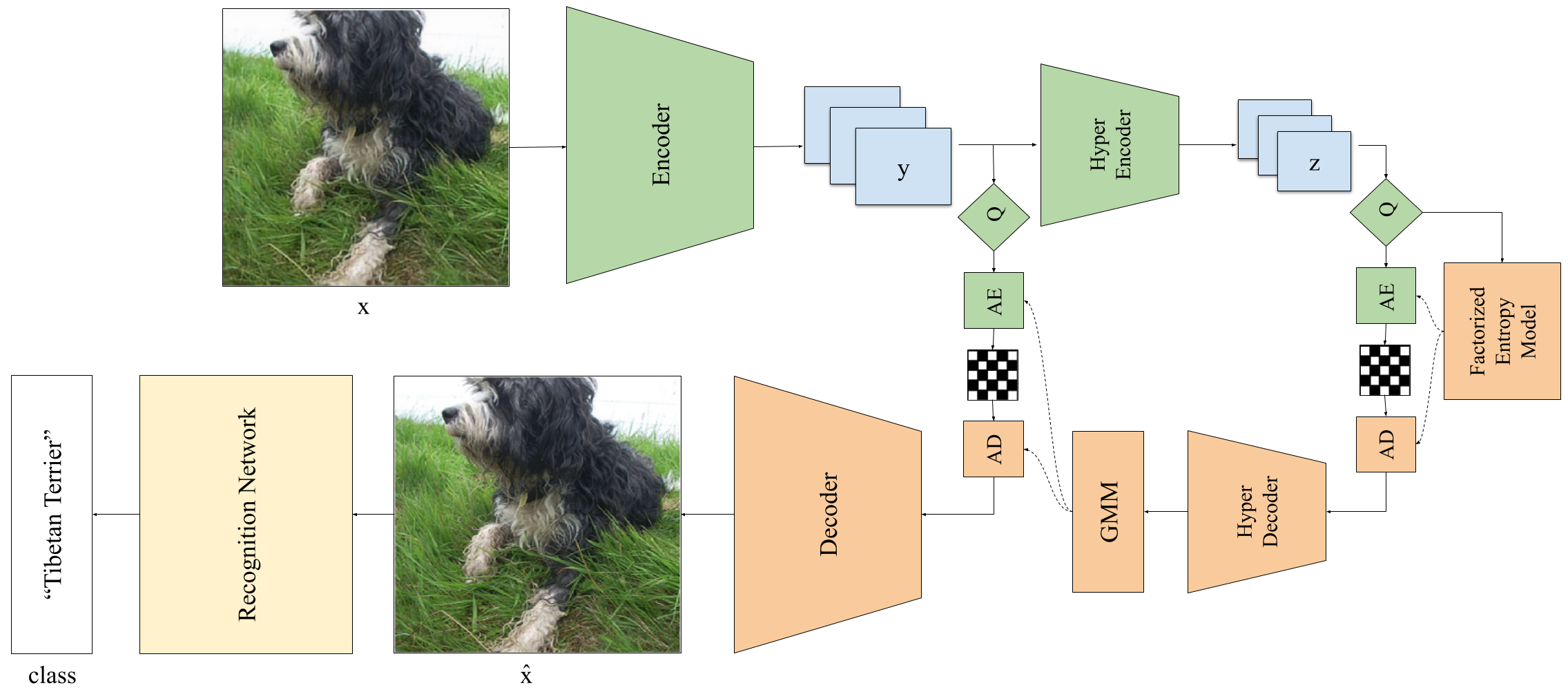}
  \caption{The joint compression-recognition architecture, where the encoders, decoders, Gaussian Mixture Model (GMM), and factorized entropy model are as in \cite{liu2020unified}. The recognition network is an EfficientNet-B0 as in \cite{DBLP:journals/corr/abs-1905-11946}. $x$ is an input image, $y$ are the latents, $z$ are the hyper-latents, $\hat{x}$ is the compressed image. AE and AD represent arithmetic encoding and decoding, respectively, and Q represents scalar rounding quantization. Dotted lines from component A to component B indicate that the outputs of A parameterize B.}
  \label{fig:architecture}
\end{figure*}

\section{Introduction}
\label{sec:intro}

Image compression, the task of reducing the storage and transmission cost of images while preserving their quality, involves three steps: transformation, quantization, and bit allocation. Traditionally, each of these steps is manually engineered and inflexible, but in recent years, learned compression methods based on convolutional neural networks have proven their ability to outperform traditional codecs by optimizing rate-distortion losses \cite{ball2016endtoend,Mentzer_2018,7906310,minnen2018joint,balle2018variational}. These convolutional neural network based methods often leverage autoencoders, architectures which repeatedly downsample input images through convolution to yield low dimensional features called latents, which can be decoded to reconstruct the image \cite{2017arXiv170300395T,Zhou_2018_CVPR_Workshops,rippel2017realtime}. 

Most deep learning methods seek optimal tradeoffs between compression efficiency and perceptual quality. As the intended consumer of the image is the human eye, compression research has focused on optimizing distortion metrics such as Peak Signal-to-Noise Ratio (PSNR) or Multiscale Structural Similarity (MS-SSIM). The bitrate, or the average number of bits required to encode a compressed image, is approximated using a model which learns to predict probability distributions over quantized latents. For a learned compression scheme, this bitrate can be approximated by the entropy of the distribution over the latents. Recent papers such as \cite{akbari2020generalized,2017arXiv170300395T,lee2019endtoend,Cheng_2020} favor Gaussian Mixture Models (GMM) with learned means, variances, and mixing weights, to model the latent distributions. Quantizing the latents is a non-differentiable operation, which presents a challenge for deep learning based approaches, but widely adopted solutions to this problem include straight-through approximation, as in \cite{bengio2013estimating}, and uniform noise approximation \cite{7906310}. Hierarchical models, pioneered in \cite{balle2018variational}, introduce a second level of compression, encoding the latents into hyper-latents which are transmitted as side information. Side information in learned compression schemes are additional bits used to improve the match between the estimated and real entropy of the latents. In GMM methods the hyperlatents are generally interpreted as the means, variances, and mixing weights for the constituent Gaussians. The bitrate of the hyper-latents must be accounted for in the loss and is usually estimated using a factorized entropy model, as introduced in \cite{minnen2018joint}.

The compression model used in our work incorporates all of these learned components: a factorized entropy model, a GMM, and a hierarchical structure. Our contribution is the addition of a task sensitivity. More and more, compressed images are consumed not by the human eye but by neural networks designed for tasks such as super-resolution or recognition. Such tasks may be sensitive to distortions not well represented by conventional distortion metrics such as PSNR, and as a result, task performance may suffer under compression by methods trained in a task-agnostic manner. Furthermore, compression methods trained using conventional metrics may be sub-optimal for a given task, allocating bits to features which, while salient for human perception, are irrelevant to task performance. 

In this work we focus on the task of recognition. Some work relevant to recognition-aware image compression has been proposed, as in \cite{Choi2018HighEC,8546064}. These methods learn spatial quantization parameter maps for compressed images based on the response strengths of feature maps from recognition networks. \cite{Sharma_2018,8803275} present methods for image enhancement driven by classification. Images are pre-transformed by convolution layers which learn to enhance the aspects of the image conducive to recognition, before being passed to recognition models. While these methods induce no explicit compression, the end-to-end nature of the training schemes are similar in spirit to what we aim to implement. In \cite{9190860}, task-specific networks are optimized with augmented losses which penalize the entropy of learned features. This encourages models to learn compressible features which can then be encoded by existing compression methods. However, no tailored compression method is jointly learned with the task. No reconstructed image is generated: rather, the task output is immediately predicted from the features, doing away with the intermediate reconstructed image. The authors are thus able to do away with the distortion term in their loss. 

\section{Proposed Approach}
\label{sec:joint}
In this paper we are interested in explicitly compressing an image and generating a reconstructed image which is passed to a recognition model. Learning the parameters of both models allows the networks to complement one another: The compression model is incentivized to allocate bits in a way which maximally preserves recognition accuracy. The recognition model is incentivized to fine tune its feature extraction layers to work efficiently with lower bitrate compressed images. As a result, we achieve higher recognition performance at lower bitrates compared to task-agnostic methods.

Most deep-learning methods optimize a problem of the form:
\begin{equation}
\label{eq:conventional_loss}
\theta^* = \argmin_{\theta} R(\hat{x}) + \lambda D(x, \hat{x})
\end{equation}
over a set of neural network parameters $\theta$, where $x$ is the original image, $\hat{x}$ is the compressed image, $R(\cdot)$ is the bitrate of the compressed image, and $D(\cdot, \cdot)$ is some distortion metric, typically mean squared error (MSE) or MS-SSIM. $\lambda$ is a Lagrange multiplier corresponding to the distortion term. We combine state-of-the-art compression and recognition models and train them jointly, learning the parameters of both models end-to-end. We optimize a three-part loss, balancing the traditional rate-distortion terms with a task-specific term added to induce a sensitivity to the recognition task. Our joint loss yields an optimization problem over the compression model\textquotesingle s parameters $\theta$ and the recognition model\textquotesingle s parameters $\phi$ of the form:
\begin{equation}
\label{eq:our_loss}
(\theta^*, \phi^*) = \argmin_{\theta, \phi} (1-\lambda) R(x) + \lambda D(x, \hat{x}) + \beta L_t(y, \hat{y})
\end{equation}
where $y$ is the true task label, $\hat{y}$ is the model's predicted task label, and $L_t$ is the task loss, in this case, cross entropy. The parameters $\lambda$ and $\beta$ allow us to control the emphasis placed on each of the constituent loss terms during training. By weighting the bitrate by $(1-\lambda)$ we couple the distortion and bitrate terms and bind $\lambda$ to the range $[0, 1]$. Note that any ratio of bitrate to distortion weighting achievable in the conventional loss with some setting $\lambda_{CL}$ is achievable in our loss with the setting $\lambda=\lambda_{CL} / (1+\lambda_{CL})$. When $\lambda$ is close to $1$ the bitrate term is severely discounted and fidelity to the original image is prized. When $\lambda$ is close to $0$ distortion is ignored and the bitrate is optimized against accuracy.

\begin{figure}[t]
\includegraphics[width=.48\textwidth]{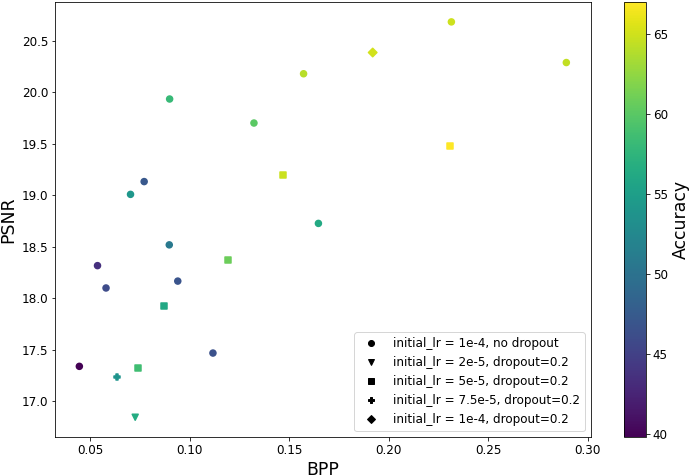}
    \caption{Bitrate in bits per pixel (BPP), accuracy, and PSNR results for our joint model at various settings of $\lambda$, $\beta$, and training parameters. Dropout was not used during training unless specified. Markers indicate training scheme, as described in legend.}
\end{figure}

\begin{figure}[t]
\includegraphics[width=.48\textwidth]{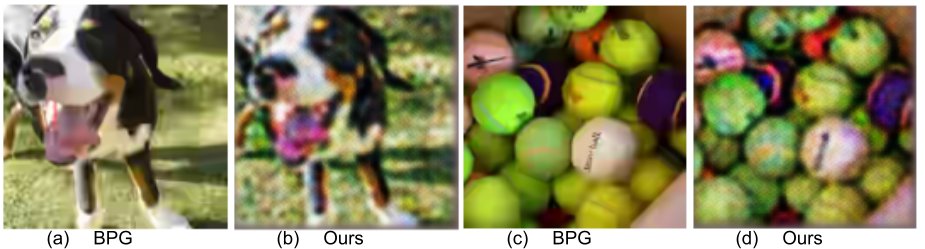}
    \caption{Sample images from BPG and our model, trained here with $\lambda=0.9, \beta=1.0$, initial learning rate of 5e-5, and dropout of 0.2. (a) has BPP=0.132, PSNR=21.55; (b) has BPP=0.119, PSNR=17.62; (c) has BPP=0.117, PSNR=26.35; (d) has BPP=0.119, PSNR=18.36.}
\end{figure}

\begin{figure*}[!htb]
\minipage[t]{0.32\textwidth}
\vspace{0pt}
  \includegraphics[width=\linewidth]{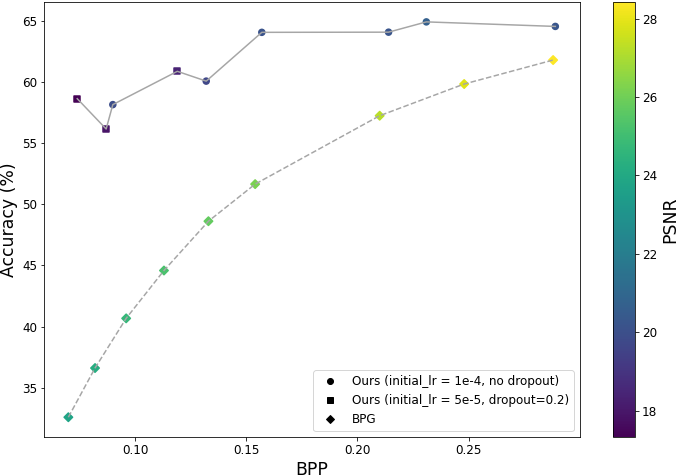}
  \caption{Comparison between the state of the art traditional codec, BPG, and our joint model, in terms of bitrate, accuracy, and PSNR. By traversing the $\lambda$, $\beta$ space we attempt to find equivalent or proximal bitrates to those achieved by BPG.}
\endminipage\hfill
\minipage[t]{0.32\textwidth}
\vspace{0pt}
  \includegraphics[width=\linewidth]{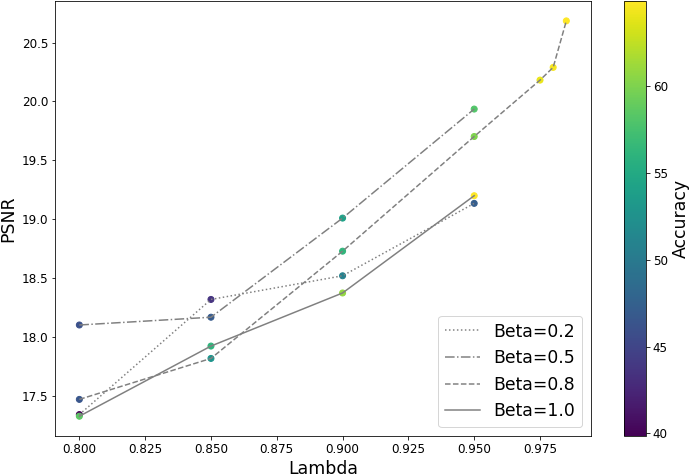}
  \caption{A demonstration of distortion control using parameters $\lambda$, $\beta$. Models with $\beta=1.0$ are trained with initial learning rate 5e-5 and dropout of 0.2; all others are trained using an initial learning rate of 1e-4 with no dropout.}
\endminipage\hfill
\minipage[t]{0.32\textwidth}%
\vspace{0pt}
  \includegraphics[width=\linewidth]{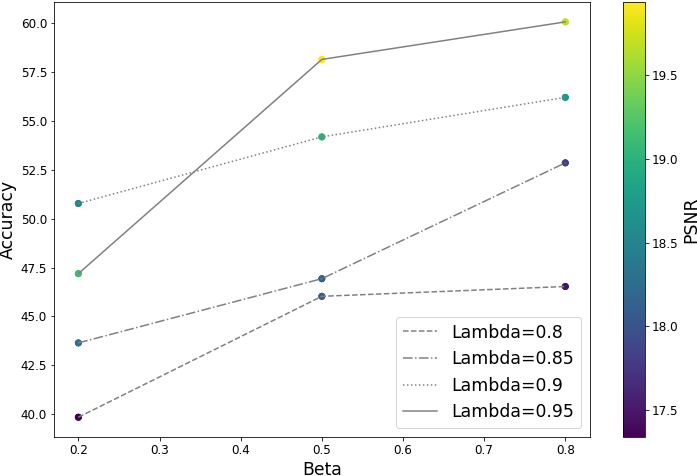}
  \caption{A demonstration of accuracy control using parameters $\lambda$, $\beta$. Models are trained using an initial learning rate of 1e-4 with no dropout.}
\endminipage
\end{figure*}

\subsection{Architecture Details}
\label{ssec:details}
Our joint architecture is illustrated in Figure 1. The compression model is based largely on the architecture from \cite{liu2020unified}, which achieves state of the art rate-distortion performance. We do away with the method\textquotesingle s proposed decoder-side enhancement module, as it largely aims to improve perceived visual quality. For the sake of simplicity we also do away with the channel attention module in the encoder and hyperencoder. As in \cite{liu2020unified} we use a GMM with two Gaussians. We also adopt the uniform noise method of quantization, adding uniform noise to the latents during training to simulate the effects of rounding in a differentiable manner.

 We add to this compression network an EfficientNet-B0 recognition model, as described in \cite{DBLP:journals/corr/abs-1905-11946}, chosen for its near state-of-the-art classification accuracy on ImageNet and low parameter count. The current state of the art on the ImageNet validation benchmark is a top-1 accuracy of 88.5\%, achieved in \cite{Touvron2020FixingTT} using a model with 480 million parameters. EfficientNet-B0 reaches a top-1 accuracy of 78.8\% but comprises only 5.3 million parameters, making its outputs usable as a heuristic for recognition accuracy without slowing down training or inference unduly.

In the compression stage, input images are passed to an encoder, which uses downsampling convolutions and Generalized Divisive Normalization (GDN) \cite{ball2015density} activation layers to yield latents -- in our case, 192 feature layers of height and width 16. These latents are passed to a hyperencoder to repeat this process and yield hyperlatents. The latents and hyperlatents are quantized. At this stage in practice they would be encoded to a bitstream using arithmetic encoding. The quantized hyperlatents are passed to the factorized entropy model, which estimates their bitrate during training, before being decoded and sent to the GMM module, which uses them to generate the means, variances, and weights for the predicted probability distributions over latents. These predicted distributions are used to estimate the training bitrate of the latents, and in practice would be used for arithmetic encoding and decoding. The quantized latents are passed to the decoder to yield the reconstructed image $\hat{x}$, which is sent to the recognition network to yield a predicted class.

\section{Experiments}
\label{sec:experiments}
We use Xavier initialization for the weights of our compression model, and initialize the EfficientNet with weights pre-trained for ImageNet classification \cite{timm}. We train our model on a random subset of 500,000 of the 1.2 million images comprising the ImageNet dataset. For validation we use the full 50,000 image validation set from the Imagenet 2012 challenge, namely ILSVRC2012. We train for 9 epochs and use MSE as the distortion metric.

Figure 2 demonstrates our model\textquotesingle s ability to reproduce the rate-distortion tradeoffs typical of compression methods. As the bitrate increases, PSNR increases and accuracy improves, a result which is indicated by the color gradient from blue to yellow. However, unlike in conventional rate-distortion curves with a one-to-one mapping between bitrates and PSNR values, our results illustrate the model\textquotesingle s ability to trade off further between PSNR and accuracy. For a given bitrate it is possible to learn models with high PSNR and low accuracy or low PSNR and high accuracy, by altering $\beta$ and training parameters such as dropout and learning rate. As in \cite{DBLP:journals/corr/abs-1905-11946} we use dropout to combat overfitting in the recognition model, adopting the suggested value of 0.2. As seen in Figure 2, using dropout significantly improves bitrate and accuracy performance. In one experiment we train two models with identical learning rate and hyperparameter settings but use no dropout for one and dropout of 0.2 for the other. We find that adding dropout decreases the bitrate from 0.289 to 0.192 BPP and increases accuracy by 0.56\%.  Additionally, through most training we adopt the initial learning rate of 1e-4, as suggested by \cite{liu2020unified} and decrease the learning rate by half during the last epoch of training. We find, however, that in the high $\lambda$ domain, e.g. $\lambda=0.999$, stability during training becomes a challenge. Lowering the learning rate to 1e-5 in such cases improves model performance. In general, performance is highly sensitive to changes in initial learning rate. Learning rate experiments included in Figure 2, where the triangle, cross, and closest square marker represent models trained identically with the exception of learning rate, demonstrate this sensitivity. 

Since our recognition model is initialized using weights pretrained on uncompressed ImageNet images, recognition performance is strongly correlated with low distortion. That is, the EfficientNet model does best when compressed input images are as close to the kinds of original, uncompressed images on which it was trained. If improvements in accuracy were due solely to lowered distortion, we would expect recognition accuracy to increase monotonically as PSNR improves. In this case, any non-joint method achieving higher PSNR at equivalent bitrates could be expected to achieve higher accuracy than our method at these points. 

However, our model demonstrates the ability to produce images with low bitrate and low PSNR, yet competitive recognition accuracy. Sample output images from our model and BPG can be seen in Figure 3; while our model at this bitrate achieves an average PSNR of 18.37 compared to BPG\textquotesingle s 25.22 on the ImageNet validation set, we achieve 16.28\% greater accuracy. This result is repeated across bitrates, as illustrated in Figure 4, which compares our results to those of BPG, the state-of-the-art traditional or engineered codec. We attempt to match the bitrates produced by BPG using $\lambda$ and $\beta$ tuning, though this targeting is fairly imprecise. We observe higher recognition accuracy at roughly equivalent bitrates, with far lower PSNR. In the low bitrate domain in particular, our method vastly outperforms BPG, achieving 26.03\% greater accuracy while producing images with PSNR lower by 6.47 on average. In this way our method makes more efficient use of allocated bits for the task at hand, optimizing for accuracy rather than visual distortion. 

Our proposed system largely reduces to EDIC, the system in \cite{liu2020unified}, when $\beta = 0$. That said, there are three differences between our system
and that of EDIC: first, we use 192 channels in our convolutions
rather than 320. Second we train on three times fewer images than
 \cite{liu2020unified}. Authors in  \cite{liu2020unified}  train their base model for 3,500,000 iterations with a
batch size of 4, exposing the model to 14,000,000 images, while
we train for 9 epochs on a dataset of 500,000 images, exposing
our model to 4,500,000 images. The training dataset in  \cite{liu2020unified}  consists of
20,745 images from Flickr and their testing set is the Kodak PhotoCD
dataset, while our training uses the aforementioned 500,000
images from ImageNet and our testing uses the full 50,000 image
ImageNet 2012 validation dataset. Third, we have not implemented
two blocks in  \cite{liu2020unified}, namely attention and decoder side enhancements,
in our model. Replicating  training in  \cite{liu2020unified} in all other ways
and running our system at $\beta = 0$, i.e. with zero weight in the loss
term for recognition accuracy, we achieve a bitrate of 0.35, PSNR
of 25.57 and recognition accuracy of 42.85\%. This PSNR is about
6.5dB less than the performance in  \cite{liu2020unified} for similar bit rates. However,
with non-zero weight for the recognition loss, e.g. $\beta = 0.2$,
we achieve a higher recognition accuracy of 66.82\%, at BPP of
0.43 and PSNR of 23.04. This demonstrates the trade off in our
work between PSNR and recognition accuracy.

Our approach to bitrate and accuracy control using $\beta$ and $\lambda$ from our loss creates a two-dimensional hyperparameter search space. For a fixed $\beta$, increasing $\lambda$ results in increased accuracy and lower distortion, and has an indeterminate effect on bitrate, as observed in Figure 5. As seen in Figure 6, we find that for a fixed $\lambda$, increasing $\beta$ results in improved recognition accuracy at the cost of a higher bitrate, and has an indeterminate effect on distortion. Within each depicted group with shared $\lambda$, we see monotonically increasing accuracy among points with identical training schemes as $\beta$ increases. We also find that changes in $\lambda$ affect model performance more than changes in $\beta$. One explanation for this is that $\lambda$ alters the model\textquotesingle s emphasis on bitrate as well as distortion while $\beta$ only indicates the emphasis on cross entropy.

\section{Conclusion and Further Work}
\label{sec:conclusion}
We present a joint approach to learned compression and recognition, training state-of-the-art models end-to-end to encourage the learning of complementary features. We demonstrate greater recognition accuracy results to those achieved by traditional methods like BPG, at equivalent bitrates. In future work we aim to extend our results to higher bitrates while remaining competitive with BPG in terms of accuracy.


{
\small
\bibliographystyle{IEEEbib}
\bibliography{refs}
}


\begin{biography}
Maxime Kawawa-Beaudan is a MS student in EECS at U.C. Berkeley advised by Professor Avideh Zakhor.

Avideh Zakhor is currently Qualcomm Chair and professor in EECS at U.C. Berkeley. Her areas of interest include theories and applications of signal, image and video processing and 3D computer vision. 
\end{biography}

\end{document}